# Characterizing the geometrical edges of nonlocal two-qubit gates

S. Balakrishnan and R. Sankaranarayanan

*Department of Physics, National Institute of Technology, Tiruchirappalli 620015, India*

Nonlocal two-qubit gates are geometrically represented by tetrahedron known as Weyl chamber within which perfect entanglers form a polyhedron. We identify that all edges of the Weyl chamber and polyhedron are formed by single parametric gates. Nonlocal attributes of these edges are characterized using entangling power and local invariants. In particular, SWAP$^{-\alpha}$ family of gates with $0 \leq \alpha \leq 1$ constitutes one edge of the Weyl chamber with SWAP$^{-1/2}$ being the only perfect entangler. Finally, optimal constructions of controlled-NOT using SWAP$^{-1/2}$ gate and gates belong to three edges of the polyhedron are presented.

## I. INTRODUCTION

There is good number of examples where quantum computation outperforms classical computation [1, 2]. Entanglement [3], nonlocal property of a quantum state having no classical counter part, is an important ingredient in quantum information processing. Since entangled states play central role in quantum algorithms, it is essential to understand the production, quantification, and manipulation of these states. Using appropriate quantum operators (gates), it is always possible to manipulate quantum states. As two-qubit gates are capable of producing entanglement, much attention has been paid to study their entangling characterization. Entangling abilities of a gate may be quantified using entangling power as introduced by Zanardi *et al* [4].

Nonlocal attributes of two-qubit gates can be characterized using local invariants [5]. Gates possessing the same local invariants are known as local equivalence class. In other words, gates belong to a local equivalence class differ only by local operations. Characterizing different local equivalence class using local invariants facilitate to reduce circuit complexity. By applying Cartan decomposition, Zhang *et al*. showed that the geometric structure of nonlocal two-qubit gates is a three torus [6]. The symmetry reduced geometry of nonlocal gates form a tetrahedron known as Weyl chamber. Each point in the Weyl chamber represents a local equivalence class of two-qubit gates. Gates that can produce maximal entanglement when acting on some separable states are called perfect entanglers and they form a polyhedron within the Weyl chamber [5, 6].

We note that while controlled unitary gates [6] and SWAP$^\alpha$ gates with $0 \leq \alpha \leq 1$ [7] form two edges of the Weyl chamber, remaining regions of the geometry are largely unexplored. The main aim of this paper is to explore other edges of the geometry using local invariants and entangling power. It is shown that SWAP$^{-\alpha}$ (inverse of SWAP$^\alpha$) form one edge of the Weyl chamber with SWAP$^{-1/2}$ being the *only* perfect entangler. Further, it is identified that *all* edges of the Weyl chamber and polyhedron are formed by single parametric gates. The local invariants and entangling power of the edges are computed and tabulated. Finally, optimal constructions of controlled-NOT gate using some of the edges are presented.

## II. PRELIMINARIES

Two unitary transformations $U, U_1 \in$ SU(4) are called locally equivalent if they differ only by local operations: $U = k_1 U_1 k_2$ where $k_1, k_2 \in$ SU(2)$\otimes$SU(2) [5]. Local equivalent class of $U$ can be associated with local invariants which are calculated as follows. An arbitrary two-qubit gate $U \in$ SU(4) can be written in the form [8, 9]

$$U = k_1 \exp\left\{\frac{i}{2}(c_1 \sigma_x^1 \sigma_x^2 + c_2 \sigma_y^1 \sigma_y^2 + c_3 \sigma_z^1 \sigma_z^2)\right\} k_2, \qquad (1)$$

where $\sigma_x, \sigma_y, \sigma_z$ are Pauli matrices. Representing $U$ in the Bell basis,

$$|\Phi^+\rangle = \frac{1}{\sqrt{2}}(|00\rangle + |11\rangle), \qquad |\Phi^-\rangle = \frac{i}{\sqrt{2}}(|01\rangle + |10\rangle),$$

$$|\Psi^+\rangle = \frac{1}{\sqrt{2}}(|01\rangle - |10\rangle), \qquad |\Psi^-\rangle = \frac{i}{\sqrt{2}}(|00\rangle - |11\rangle)$$

as $U_B = Q^\dagger U Q$, with

$$Q = \frac{1}{\sqrt{2}} \begin{pmatrix} 1 & 0 & 0 & i \\ 0 & i & 1 & 0 \\ 0 & i & -1 & 0 \\ 1 & 0 & 0 & -i \end{pmatrix},$$

the local invariants are defined as [5, 6]

$$G_1 = \frac{tr^2[M(U)]}{16 \det(U)}, \tag{2a}$$

$$G_2 = \frac{tr^2[M(U)] - tr[M^2(U)]}{4 \det(U)}, \tag{2b}$$

where $M(U) = U_B^T U_B$. Local invariants and a point $[c_1, c_2, c_3]$ on a 3-torus geometric structure are related as [10]

$$G_1 = \frac{1}{4}\left[e^{-ic_3}\cos(c_1 - c_2) + e^{ic_3}\cos(c_1 + c_2)\right]^2, \tag{3a}$$

$$G_2 = \cos(2c_1) + \cos(2c_2) + \cos(2c_3). \tag{3b}$$

From these relations, geometrical point on a three torus corresponding to a local equivalence class of gates can be identified. The symmetry reduced three torus takes the form of tetrahedron (Weyl chamber) as shown in Fig.1.

A two-qubit gate is called a perfect entangler if it produces a maximally entangled state for some separable input state. The theorem for a perfect entangler is the following: two-qubit gate $U$ is a perfect entangler if and only if the convex hull of the eigenvalues of $M(U)$ contains zero [5, 6]. Alternatively, if the coordinates satisfy the condition

$$\frac{\pi}{2} \le c_i + c_k \le c_i + c_j + \frac{\pi}{2} \le \pi \quad \text{or} \quad \frac{3\pi}{2} \le c_i + c_k \le c_i + c_j + \frac{\pi}{2} \le 2\pi, \tag{4}$$

where $(i, j, k)$ is a permutation of (1,2,3), then the corresponding two-qubit gate is a perfect entangler. Thus the perfect entangler nature of a given gate $U$ can be ascertained from the corresponding geometric representation. Perfect entanglers within the Weyl chamber constitute a polyhedron. In fact, exactly half of the nonlocal two-qubit gates are perfect entanglers [6].

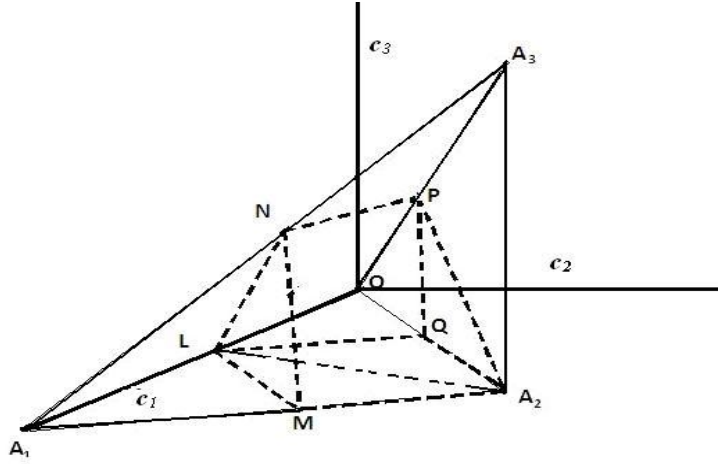

FIG. 1. Tetrahedron $OA_1A_2A_3$, the geometrical representation of nonlocal two-qubit gates, is referred as Weyl chamber. Polyhedron $LMNPQA_2$ (shown in dashed lines) corresponds to the perfect entanglers. The thick lines represent the $c_1$, $c_2$ and $c_3$ axes of the Weyl chamber. The points $L$, $M$, $N$, $P$, and $Q$ are midpoints of the tetrahedron edges $OA_1$, $A_2A_1$, $A_3A_1$, $OA_3$, and $OA_2$ respectively. Line $LA_2$ represents special perfect entanglers. The points $L = [\pi/2, 0, 0]$, $A_3 = [\pi/2, \pi/2, \pi/2]$, $P = [\pi/4, \pi/4, \pi/4]$ and $N = [3\pi/4, \pi/4, \pi/4]$ correspond to CNOT, SWAP, $\text{SWAP}^{1/2}$, and $\text{SWAP}^{-1/2}$ gates respectively.

Entangling capability of a unitary quantum gate $U$ can be quantified by entangling power [4]. For a unitary operator $U \in U(4)$ the entangling power is defined as

$$e_p(U) = \overline{[E(U|\psi_1\rangle \otimes |\psi_2\rangle)]}_{|\psi_1\rangle \otimes |\psi_2\rangle} \tag{5}$$

where the overbar denotes the average over all product states distributed uniformly in the state space. In the above formula $E$ is the linear entropy of entanglement measure defined as

$$E(|\psi\rangle_{AB}) = 1 - tr(\rho^2_{A(B)}) \tag{6}$$

where $\rho_{A(B)} = tr_{B(A)}(|\psi\rangle_{AB}\langle\psi|)$ is the reduced density matrix of system $A(B)$. Various properties of $e_p$ can be found elsewhere [4,11,12,13]. In terms of coordinates, entangling power of a unitary operator can be written as [10]

$$e_p(U) = \frac{1}{18}[3 - (\cos 2c_1 \cos 2c_2 + \cos 2c_2 \cos 2c_3 + \cos 2c_3 \cos 2c_1)]. \tag{7}$$

## III. SWAP$^{-\alpha}$ FAMILY OF GATES

General form of SWAP$^{-\alpha}$ is given by

$$\text{SWAP}^{-\alpha} = \begin{pmatrix} 1 & 0 & 0 & 0 \\ 0 & \dfrac{1+\exp(-i\pi\alpha)}{2} & \dfrac{1-\exp(-i\pi\alpha)}{2} & 0 \\ 0 & \dfrac{1-\exp(-i\pi\alpha)}{2} & \dfrac{1+\exp(-i\pi\alpha)}{2} & 0 \\ 0 & 0 & 0 & 1 \end{pmatrix} \quad (8)$$

with $0 \leq \alpha \leq 1$. This family of gates is the inverse of SWAP$^{\alpha}$ introduced in Ref. [7, 14]. Using Eqs. (2a) and (2b) the local invariants are obtained as

$$G_1 = \frac{1}{16}\left[9\exp(i\pi\alpha) + \exp(-i3\pi\alpha) + 6\exp(-i\pi\alpha)\right], \quad (9a)$$

$$G_2 = 3\cos(\pi\alpha). \quad (9b)$$

Subsequently the corresponding geometrical points can be evaluated from Eqs. (3a) and (3b) as

$$[c_1, c_2, c_3] = \left[\pi - \frac{\pi\alpha}{2}, \frac{\pi\alpha}{2}, \frac{\pi\alpha}{2}\right], \quad (10)$$

which represents the edge $A_1A_3$ of the Weyl chamber. For these points, condition (4) becomes

$$\frac{1}{2} \leq 1 \leq \frac{2\alpha+1}{2} \leq 1 \quad \text{or} \quad \frac{3}{2} \leq 1 \leq \frac{2\alpha+1}{2} \leq 2.$$

It is easy to check that the first inequality is satisfied only for $\alpha = 1/2$ and second inequality is invalid. Hence one can infer that SWAP$^{-1/2}$ is the *only* perfect entangler and the corresponding point is $N = [3\pi/4, \pi/4, \pi/4]$. This is clearly seen in Fig.1 that the only point $N$ of the edge $A_1A_3$ belongs to the polyhedron.

The entangling power is obtained from Eq. (7) as

$$e_p\left(SWAP^{-\alpha}\right) = \frac{1}{12}\left[1 - \cos(2\pi\alpha)\right]. \quad (11)$$

which is the same as that of SWAP$^{\alpha}$ [7, 14]. Since SWAP$^{\alpha}$ and SWAP$^{-\alpha}$ are inverse to each other, they possess the same entangling power. It is worth mentioning that $e_p(SWAP^{-1/2}) = 1/6$ and it is the maximum value in the SWAP$^{-\alpha}$ family of gates. Further,

CNOT can also be constructed using two SWAP$^{-1/2}$ gates as shown below:

$$CNOT \equiv SWAP^{-1/2} (\sigma_x \otimes \sigma_y) SWAP^{-1/2}$$
$$CNOT \equiv SWAP^{-1/2} (\sigma_x \otimes \sigma_z) SWAP^{-1/2}. \quad (12)$$

The above shown equivalence can be verified using local invariants (for CNOT $G_1 = 0$ and $G_2 = 1$).

## IV. CHARACTERIZATION OF THE EDGES

In this section we characterize six edges of the Weyl chamber and nine edges of the polyhedron (perfect entanglers) using entangling power and local invariants computed from geometrical points. Using Eqs. (3) and (7), local invariants and entangling power of the edges are tabulated in Tables I and II.

TABLE I. Entangling power and local invariants of Weyl chamber edges.

| Edge | $[c_1, c_2, c_3]$ | Range of the parameter | $e_p$ | $G_1$ | $G_2$ |
|---|---|---|---|---|---|
| $OA_1$ | $[\theta, 0, 0]$ | $0 \leq \theta \leq \pi$ | $\frac{1}{9}[1 - \cos(2\theta)]$ | $\cos^2(\theta)$ | $2\cos^2(\theta) + 1$ |
| $OA_2$ | $[\theta, \theta, 0]$ | $0 \leq \theta \leq \pi/2$ | $\frac{1}{18}[3 - \cos^2(2\theta) - 2\cos(2\theta)]$ | $\frac{1}{4}[1 + \cos(2\theta)]^2$ | $1 + 2\cos(2\theta)$ |
| $A_2A_1$ | $\left[\frac{\pi}{2} + \phi, \frac{\pi}{2} - \phi, 0\right]$ | $0 \leq \phi \leq \pi/2$ | $\frac{1}{18}[3 - \cos^2(2\phi) + 2\cos(2\phi)]$ | $\frac{1}{4}[\cos(2\phi) - 1]^2$ | $1 - 2\cos(2\phi)$ |
| $A_2A_3$ | $\left[\frac{\pi}{2}, \frac{\pi}{2}, \phi\right]$ | $0 \leq \phi \leq \pi/2$ | $\frac{1}{9}[1 + \cos(2\phi)]$ | $-\sin^2(\phi)$ | $\cos(2\phi) - 2$ |
| $OA_3$ | $\left[\frac{\pi\alpha}{2}, \frac{\pi\alpha}{2}, \frac{\pi\alpha}{2}\right]$ | $0 \leq \alpha \leq 1$ | $\frac{1}{12}[1 - \cos(2\pi\alpha)]$ | $\frac{1}{16}\left[9e^{-i\pi\alpha} + e^{3i\pi\alpha} + 6e^{i\pi\alpha}\right]$ | $3\cos(\pi\alpha)$ |
| $A_1A_3$ | $\left[\pi - \frac{\pi\alpha}{2}, \frac{\pi\alpha}{2}, \frac{\pi\alpha}{2}\right]$ | $0 \leq \alpha \leq 1$ | $\frac{1}{12}[1 - \cos(2\pi\alpha)]$ | $\frac{1}{16}\left[9e^{i\pi\alpha} + e^{-3i\pi\alpha} + 6e^{-i\pi\alpha}\right]$ | $3\cos(\pi\alpha)$ |

While SWAP$^\alpha$ family of gates lie along the edge $OA_3$ [7], their inverse form the edge $A_1A_3$ as shown in the earlier section. Line $OL \in OA_1$ consists of controlled unitary gates with CNOT being represented by the point $L = [\pi/2, 0, 0]$. In $c_1c_2$ – plane, base areas $LA_2A_1$ and $LA_2O$ are mirror images (locally equivalent) to each other with respect to the line $LA_2$. This line is identified as special perfect entanglers (SPE) possessing the maximum entangling power of 2/9 [10]. Figure 2 shows the entangling power of $OA_1$ and $OA_2$ with respect to their parameter. Similarly, the entangling power of $A_2A_1$ and $A_2A_3$ (perpendicular to $c_1c_2$ - plane) is shown in Fig.3. From the plots, it is clear that the entangling power of these edges varies monotonically.

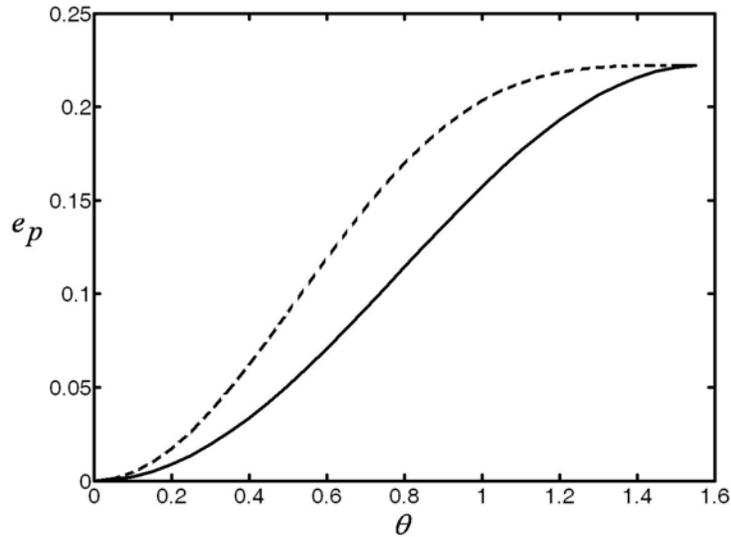

FIG. 2. Entangling power of $OL$ (solid line) and $OA_2$ (dashed line). The other half of $OA_1$ is not shown due to symmetry in the base.

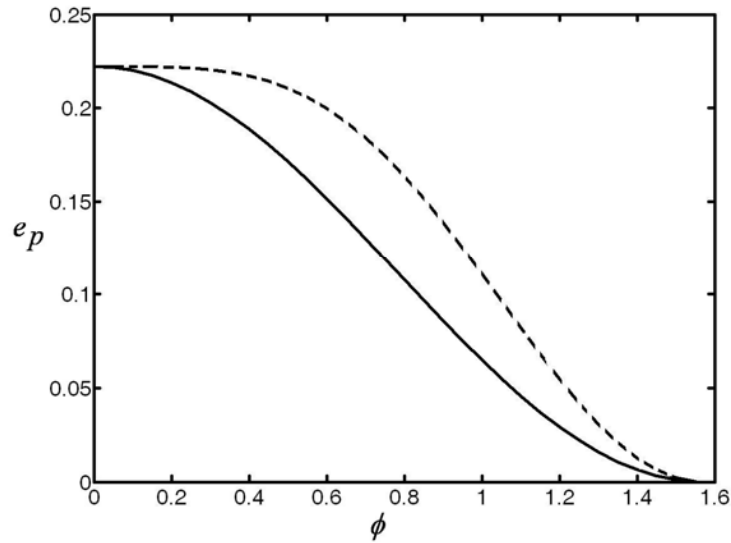

FIG. 3. Entangling power of $A_2A_3$ (solid line) and $A_2A_1$ (dashed line).

TABLE II. Entangling power and local invariants of edges of the polyhedron.

| Edge | $[c_1, c_2, c_3]$ | Range of the parameter | $e_p$ | $G_1$ | $G_2$ |
|---|---|---|---|---|---|
| LQ | $\left[\frac{\pi}{2}-\theta, \theta, 0\right]$ | $0 \leq \theta \leq \pi/4$ | $\frac{1}{18}[3+\cos^2(2\theta)]$ | $\frac{1}{4}[\sin(2\theta)]^2$ | 1 |
| LM | $\left[\frac{\pi}{2}+\theta, \theta, 0\right]$ | $0 \leq \theta \leq \pi/4$ | $\frac{1}{18}[3+\cos^2(2\theta)]$ | $\frac{1}{4}[\sin(2\theta)]^2$ | 1 |
| $A_2M$ | $\left[\frac{\pi}{2}+\phi, \frac{\pi}{2}-\phi, 0\right]$ | $0 \leq \phi \leq \pi/4$ | $\frac{1}{18}[3-\cos^2(2\phi)+2\cos(2\phi)]$ | $\frac{1}{4}[1-\cos(2\phi)]^2$ | $1-2\cos(2\phi)$ |
| $A_2Q$ | $\left[\frac{\pi}{2}-\phi, \frac{\pi}{2}-\phi, 0\right]$ | $0 \leq \phi \leq \pi/4$ | $\frac{1}{18}[3-\cos^2(2\phi)+2\cos(2\phi)]$ | $\frac{1}{4}[1-\cos(2\phi)]^2$ | $1-2\cos(2\phi)$ |
| QP | $\left[\frac{\pi}{4}, \frac{\pi}{4}, \eta\right]$ | $0 \leq \eta \leq \pi/4$ | $\frac{1}{6}$ | $\frac{1}{4}[e^{-i2\eta}]$ | $\cos(2\eta)$ |
| MN | $\left[\frac{3\pi}{4}, \frac{\pi}{4}, \eta\right]$ | $0 \leq \eta \leq \pi/4$ | $\frac{1}{6}$ | $\frac{1}{4}[e^{i2\eta}]$ | $\cos(2\eta)$ |
| PN | $\left[\frac{\pi}{4}+\eta, \frac{\pi}{4}, \frac{\pi}{4}\right]$ | $0 \leq \eta \leq \pi/2$ | $\frac{1}{6}$ | $-\frac{i}{4}[e^{-i2\eta}]$ | $-\sin(2\eta)$ |
| LN | $\left[\frac{\pi}{2}+\theta, \theta, \theta\right]$ | $0 \leq \theta \leq \pi/4$ | $\frac{1}{18}[3+\cos^2(2\theta)]$ | $\frac{1}{4}[e^{i\theta}(\sin(2\theta))]^2$ | $\cos(2\theta)$ |
| $A_2P$ | $\left[\frac{\pi}{2}-\theta, \frac{\pi}{2}-\theta, \theta\right]$ | $0 \leq \theta \leq \pi/4$ | $\frac{1}{18}[3+\cos^2(2\theta)]$ | $-\frac{1}{4}[e^{i\theta}(\sin(2\theta))]^2$ | $-\cos(2\theta)$ |

Perfect entanglers within the Weyl chamber form a polyhedron $LMNPQA_2$, and the nonlocal characteristics of its edges are shown in Table II. Four edges of polyhedron lying in $c_1c_2$ - plane are $LQ$, $LM$, $A_2Q$, and $A_2M$. Since $LQ$ ($A_2Q$) and $LM$ ($A_2M$) are locally equivalent to each other, they possess same local invariants and entangling power. We may note that the edge $LQ(LM)$ is parallel to $A_2M(A_2Q)$. The edge $QP$ and $MN$ are perpendicular to $c_1c_2$ - plane and parallel to each other. The edge $PN$ is parallel to $c_1$ at $c_3 = \pi/4$ plane. It is worth noting that the edges $QP$, $MN$, and $PN$ possess the same entangling power of 1/6, though they are not locally equivalent to each other. We also note that the edges $LQ$, $LN$, and $A_2P$ of different locally equivalent classes assume same expression for their entangling power. The entangling power of *LQ* and *A₂M* varies monotonically with the parameter, as shown in Fig. 4.

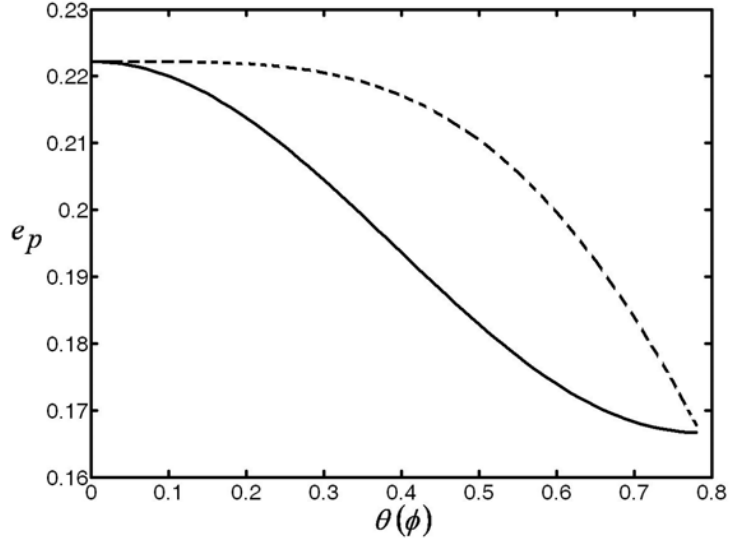

FIG. 4. Entangling power of $LQ$ (solid line) and $A_2 M$ (dashed line).

## V. CIRCUIT EQUIVALENCE OF CNOT

Universality is one of the central issues in quantum circuit complexity. Since CNOT is believed to be a universal quantum gate, at least for a wide class of controlled quantum logics [15], it is natural to look for CNOT construction from other gates. It is well known that CNOT can be constructed using two SWAP$^{1/2}$ gates [16]. Another possibility of CNOT simulation is using two gates belong to the line *LA₂*, namely SPE [10]. In this section we present new possibilities of CNOT construction using three edges of the polyhedron, namely *QP*, *MN*, and *PN*.

It is to be mentioned that the Weyl chamber can be described by $\pi/2 \geq c_1 \geq c_2 \geq |c_3|$ in the $(c_1, c_2, c_3)$ space. By incorporating this inequality, Eq. (1) can be written in the standard computational basis as [10]

$$U = \begin{pmatrix} e^{-ic_3/2} c^- & 0 & 0 & -ie^{-ic_3/2} s^- \\ 0 & e^{ic_3/2} c^+ & -ie^{ic_3/2} s^+ & 0 \\ 0 & -ie^{ic_3/2} s^+ & e^{ic_3/2} c^+ & 0 \\ -ie^{-ic_3/2} s^- & 0 & 0 & e^{-ic_3/2} c^- \end{pmatrix}. \quad (13)$$

where $c^{\pm} = \cos[(c_1 \pm c_2)/2]$ and $s^{\pm} = \sin[(c_1 \pm c_2)/2]$. Denoting $A_{QP}$ as matrix form of the edge *QP*, it is easy to verify the equivalence

$$CNOT \equiv A_{QP}(I \otimes \sigma_x)A_{QP},$$
$$CNOT \equiv A_{QP}(I \otimes \sigma_y)A_{QP}, \qquad (14)$$

using local invariants. Similarly, CNOT can be constructed from the edge *MN* as

$$CNOT \equiv A_{MN}(\sigma_x \otimes \sigma_z)A_{MN},$$
$$CNOT \equiv A_{MN}(\sigma_y \otimes \sigma_z)A_{MN}, \qquad (15)$$

where $A_{MN}$ represents matrix form of the edge *MN*. Further, such a construction is also possible using the edge *PN*:

$$CNOT \equiv A_{PN}(\sigma_x \otimes \sigma_z)A_{PN} \qquad (16)$$

where $A_{PN}$ is the matrix form of *PN*.

From Table II we note that the edges *QP*, *MN*, and *PN* are single parametric gates with entangling power 1/6. Interestingly, local invariants of all the above constructions are independent of the parameter $\eta$. In other words, *all* the gates belong to the three edges are capable of generating CNOT class. Further, the given constructions are *optimal* in the sense that the number of nonlocal gates utilized is only two. In this context, it is worth identifying suitable physical models for realizing these edges. For example, anisotropic Heisenberg spin system may realize some of the gates mentioned above. This deserves a detailed investigation which is currently underway.

## VI. SUMMARY

In this paper we have explored the geometry of nonlocal two-qubit gates (Weyl chamber). In particular, edges of the Weyl chamber and polyhedron are characterized by entangling power and local invariants. It is identified that SWAP$^\alpha$ gates and their inverse (SWAP$^{-\alpha}$) constitute two edges of the Weyl chamber. Further, it is found that the gate with $\alpha = 1/2$ is the *only* perfect entangler in the above family possessing maximum entangling power of 1/6. From the geometry of nonlocal gates, it is observed that *all* the six edges of the Weyl chamber are formed by *single* parametric gates. The entangling power and local invariants of these edges are shown in Table I. Leaving two edges formed by SWAP$^\alpha$ gates and their inverse, entangling power of the other four edges is shown to be monotonic function of the parameter.

Similarly, *all* edges of the polyhedron (perfect entanglers) are also formed by *single* parametric gates, whose entangling power and local invariants are tabulated in Table II. Interestingly, three edges of the perfect entanglers belong to different locally equivalent class possess the *same* entangling power of 1/6. Finally, numerous possibilities of optimal construction of CNOT using these edges are also shown.

---